\documentclass[aps,twocolumn,floatfix]{revtex4}
\usepackage{epsfig}

\def \S{{\cal S}}

\begin{document}

\title{Irradiated Graphene Nanostructures as Nanoscopic Pressure Cells}

\author{Michael Zaiser and Stefan Chartier}

\affiliation{The University of Edinburgh, School of Engineering, Institute for Materials and Processes,
The King's Buildings, Sanderson Building, Edinburgh EH11DT, UK}
\date{\today}

\begin{abstract}
High-dose irradiation of nanostructures consisting of multiple graphene shells, such as spherical 'carbon onions' (CO) or cylindrical multi-walled carbon nanotubes (MWNT), induces shell shrinkage by bond reconstruction around irradiation-induced defects. This leads to build-up of internal stresses and to extreme pressures acting on encapsulated materials or on the carbon cores of the nanostructures. We formulate a model which relates the build-up of internal stresses to the point defect dynamics in graphene shells. Calculations are performed for the special case of irradiated CO. The results are in good agreement with experimental findings.
\end{abstract}

\maketitle

Carbon onions (CO) consisting of concentric spherical graphene shells form under intense irradiation of graphitic or amorphous carbon with electrons \cite{banhart96,banhart96a,banhart98} or ions \cite{wesolowski97}. At temperatures above 600 K, in-situ annealing of radiation defects leads to coherent graphene shells which contract during irradiation as manifested by decreasing shell spacing (Figure 1) \cite{banhart96}. This process may create sufficiently high pressures in the CO cores to nucleate diamond crystallites \cite{redlich98} which continue to grow under irradiation \cite{zaiser97}. Studies of irradiated CO with encapsulated metal crystallites provide additional evidence for large pressures which lead to a substantial shift in the melting temperature of the encapsulated material \cite{banhart03}. Hydrostatic pressures above 20GPa were demonstrated by determining the lattice constant of Au nanoparticles encapsulated in irradiated CO \cite{sun08}. Pressurization also occurs in MWNT. Irradiation-induced MWNT contraction was observed to reduce the lattice constants of encapsulated metals and carbides, indicating pressures up to 40 GPa, and to induce plastic deformation at stresses close to the theoretical shear stress \cite{sun06}. 
\begin{figure}[b]
\begin{minipage}{8.5cm}
\centerline{\includegraphics[height=6cm]{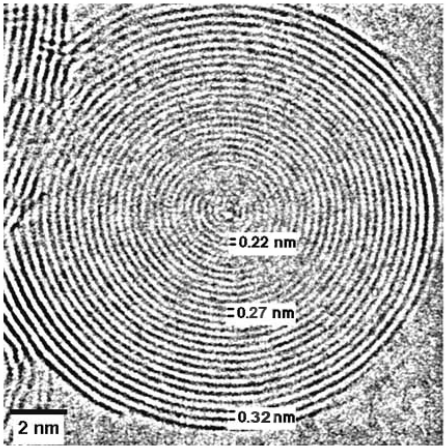}}
\caption{CO during 1.2MeV e$^-$ irradiation 
 at 1000 K; coherent shells with decreased spacing towards the core 
\protect\cite{banhart98}.}
\end{minipage}
\end{figure}
Shell shrinkage has been attributed to bond re-arrangement around irradiation-induced point defects. Simulations indicate  reconstruction of di-vacancies which leads to saturation of dangling bonds and concomitant 'loss' of unoccupied sites in the $sp^2$ network \cite{bates98,sammalkorpi04}. In single-walled nanotubes this decreases the average nanotube radius by an amount that is proportional to the di-vacancy concentration \cite{sun06}. In multi-shell structures such as CO and MWNT the situation is more complex owing to the migration and reactions of point defects within multiple shells \cite{banhart98,banhart99}. Existing rate equation models for the point-defect dynamics in irradiated CO \cite{sigle97} do not account for the loss of lattice sites by di-vacancy re-arrangement. In the present work, we formulate a model of the dynamics of point-defects in closed-shell graphene nanostructures which for the first time accounts for the di-vacancy reconstruction mechanism, the concomitant geometrical changes, and the evolution of internal stresses. 

We consider a spherical or cylindrical graphene nanostructure consisting of $M$ concentric shells ${\cal S}_i$, $1 \le i \le M$. The outermost shell is labeled ${\cal S}_1$. $N_i$ is the number of sp$^2$ bonded sites on ${\cal S}_i$. (Note that the shells may not possess perfect Graphene structure but contain pentagons, Stone-Wales defects \cite{stone86}, or more complex bonding patterns.) The stress-free radius of $\S_i$ is $r_i = [N_i A^*/(4 \pi)]^{1/2}$ for CO and $r_i = N_i A^*/(\pi l)$ for MWNT where $A^*$ is the average area per atom which we approximate by the value for planar graphene and $l \gg r_i$ the MWNT length. The spacing between shells $i$ and $i+1$ is $d_i := r_i - r_{i+1}$. The number of vacant lattice sites on $\S_i$ is $n_i^{\rm V}$, and the number of interstitials contained between $\S_i$ and $\S_{i+1}$ is $n_i^{\rm I}$. These interstitials diffuse two-dimensionally on an 'interstitial shell' of area $A_i$ ($A_i = \pi (r_i + r_{i+1})^2$ for CO). Diffusion of vacancies is considered negligible because of their significantly higher migration energy \cite{banhart99}. 

We envisage the following processes: (i) Irradiation displaces atoms at rate $\Phi$ from sp$^2$ bonded sites. Displacement of an atom from $\S_i$ creates a vacancy on this shell and, with equal probability, an interstitial either between $\S_i$ and $\S_{i+1}$ ($n_i^{\rm I} \to n_i^{\rm I} + 1$) or between $\S_i$ and $\S_{i-1}$ ($n_{i-1}^{\rm I} \to n_{i-1}^{\rm I} + 1$). Accounting for vacant lattice sites, the total rate for this process is $\Phi (N_i - n^{\rm V}_i)$. (ii) Recombination of interstitials with vacancies on adjacent shells is modelled as a bimolecular diffusion-controlled reaction where the reaction constant $K \approx A^* \nu_0 \exp [-E_{\rm M}^{\rm I}/(k_{\rm B}T)]$ approximately equals the diffusion rate of interstitials between their enclosing shells, $\nu_0 \approx 10^{13}$ s$^{-1}$ is a frequency of the order of the Debye frequency and $E_{\rm M}$ the interstitial migration enthalpy. Stress gradients affect the recombination rates: If there is exactly one vacancy on $\S_i$ and one on $\S_{i+1}$, then in absence of internal stresses an interstitial diffusing between these shells has equal probabilities $P_{i, i} = P_{i, i+1} = 1/2$ to recombine with either vacancy. If the graphene shells experience different in-plane stresses $\sigma_i$, on the other hand, we assume that this leads to preferential recombination of interstitials in the direction of increased tensile/reduced compressive stress. The modified recombination probabilities follow from 
\begin{equation}
P_{i\to i} + P_{i, i+1} = 1 \;,\quad
\frac{P_{i\to i}}{P_{i, i+1}} = \exp 
\left[\frac{(\sigma_i - \sigma_{i+1}) V}{k_{\rm B}T}\right],
\end{equation} 
where $V = \partial E^{\rm V}/\partial \sigma$ characterizes the dependence of the vacancy energy $E^{\rm V}$ in graphene on in-plane tensile or compressive stresses. Thus, creation and recombination of interstitials may lead to a net flux of matter in the direction of decreasing compressive/increasing tensile stresses. (iii) Shells shrink when irradiation creates di-vacancies
which reconstruct to saturate the dangling bonds, resulting in two pentagons and one octagon in the bond network (see e.g. \cite{bates98,sammalkorpi04}). The resulting defects do not recombine with interstitials as the energy of a reconstructed divacancy is {\em less} than that of a single vacancy \cite{krasheninnikov06} - we may thus think of di-vacancy formation on ${\cal S}_i$ as the removal of two sites from the sp$^2$ network, $N_i \to N_i -2$, and of two vacancies, $n_i^{\rm V} \to n_i^{\rm V} -2$. For simplicity, we assume that $A^*$ is not changed by this process which thus reduces the shell surface by $2A^*$ and the mean shell radius by $A^*/(4 \pi r_i)$ for CO.  With vacancy diffusion assumed negligible, di-vacancy creation is governed by the displacement of nearest neighbors of pre-existing vacancies, and occurs at rate $3 \Phi n^{\rm V}_i$. 

Combining processes (i) to (iii), we arrive at coupled rate equations for the vacancy and interstitial numbers:
\begin{eqnarray}
\partial_t n_i^{\rm V} &=& \Phi(N_i - 7 n_i^{\rm V}) \nonumber\\
&-& K n_i^{\rm V}
\left[P_{i, i}\frac{n_i^{\rm I}}{A_i} + P_{i-1, i}\frac{n_{i-1}^{\rm I}}{A_{i-1}} \right]\;,
\label{ratevac}
\\
\partial_t n_i^{\rm I} &=& \Phi(N_i + N_{i+1}- n_i^{\rm V}) - n_{i+1}^{\rm V})/2
\nonumber\\
&-& K n_i^{\rm I}
\left[P_{i, i}\frac{n_i^{\rm V}}{A_i} + P_{i, i+1} \frac{n_{i+1}^{\rm V}}{A_i}\right]\;.
\label{rateint}
\end{eqnarray}
These equations need to be modified in the core ($i=M$) where $P_{M, M} = 1$ and $N_{M+1}= n_{M+1}^{\rm V}=0$, and on the outermost shell $\S_1$ where $P_{0,1} =0$ as atoms that are displaced outwards do not recombine with vacancies but leave the system. The rate of contraction of ${\cal S}_i$ due to di-vacancy creation is given by $\partial_t r_i = - 3 \Phi n_i^{\rm V} A^*/(4\pi r_i)$ for CO. 
\begin{figure*}[t]
\centerline{
\psfig{figure=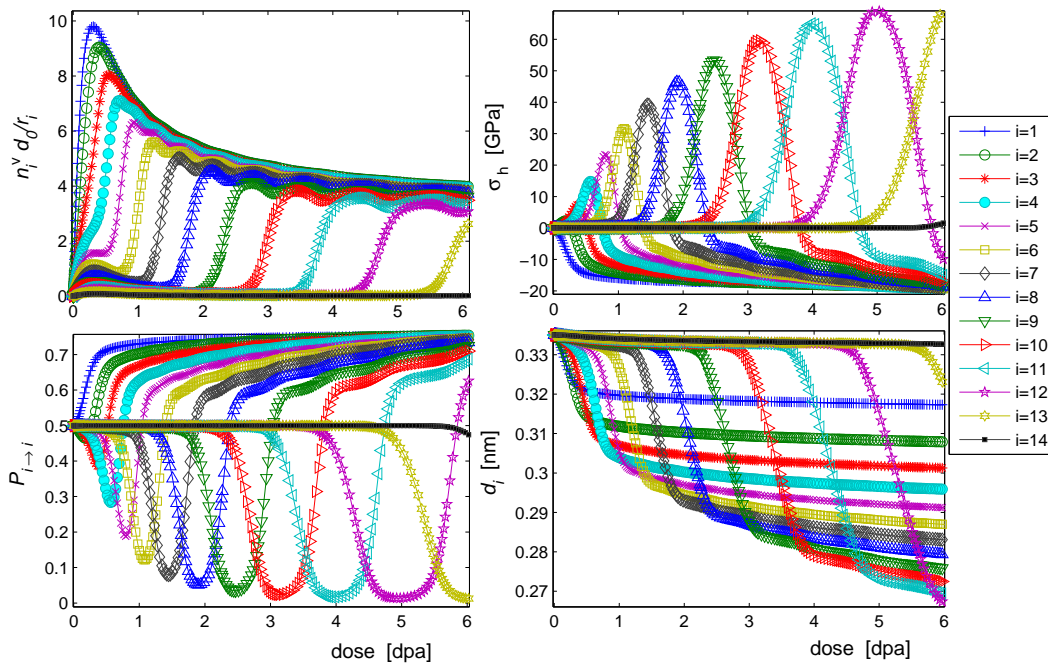,width=0.8\textwidth}
}\caption{Evolution of the vacancy number $n_i^{\rm V}$ (multiplied with $d_0/r_i$), hoop stress $\sigma_{\rm h}$, recombination probability $P_{i\to i}$, and shell spacing $d_i$ during the first stage of irradiation of a CO with 16 shells; $\Phi = 0.01$ dpa/s, $T = 570 K$, other parameters see text.}
\label{Fig:shell_core}
\end{figure*}

To obtain a closed description, the geometry needs to be related to the internal stresses acting in the structure. To this end, we consider the inner and outer pressure acting on ${\cal S}_i$ because of Van der Waals interactions with the neighbouring shells. If there is a difference between inner and outer pressure, force equilibrium requires this to be balanced by a hoop stress. Considering ${\cal S}_i$ as a thin shell of radius $r_i$ and thickness $d_i$, the hoop stress is given by $\sigma_{\rm h} = (f_{i}-f_{i-1})r_i/(\alpha d_i)$ where $f_i$ is the van der Waals force per unit area acting on the shell from inside due to its interaction with shell $i+1$, $\alpha =1$ for MWNT and $\alpha=2$ for CO. To evaluate the forces, we refer to experimental data for planar graphite. In a third-order polynomial approximation, the force per unit area acting on ${\cal S}_i$ from inside is approximated by $f_{i} = c_{33}[\varepsilon_i^{\rm c} + K_1 (\varepsilon_i^{\rm c})^2 + K_2 [(\varepsilon_i^{\rm c})^3]$ where $\varepsilon_i^{\rm c} = (d_i/d_0) - 1$ and we use for the elastic constant $c_{33}$ the value for bulk graphite, $c_{33} \approx 36$ GPa \cite{michel08}. $d_0 \approx 0.335$nm is the shell spacing in absence of internal stresses which we take to equal the $c$-axis lattice spacing of bulk graphite. The polynomial coefficients are determined by fitting to experimental pressure-compression data \cite{lynch66}, yielding $K_1 \approx 0$, $K_2 \approx 130$, i.e., the third-order non-linear corrections prevail; they are substantial even for moderate degrees of compression. The boundary condition at the outer surface of the nanostructure is $p_{0,1}=0$. To evaluate $d_i$ and $r_i$, we consider the shells as rigid and approximate the shell radii by their stress-free values. This is justified by the extreme elastic anisotropy of graphitic structures, as the in-plane elastic modulus of the graphene layers exceed the modulus in the perpendicular direction by a factor of 30. 

With an interstitial migration energy $E_{\rm I}^{\rm M} = 0.8$ eV \cite{banhart99} and typical irradiation rates of $\Phi \approx 10^{-4}\dots 1$ dpa/s, the model is applicable to temperatures $T > 230$ K$ - 310$ K, below which defect concentrations become so large that the shells lose coherency. We thus exclusively consider irradiation at above-ambient temperatures and explore the experimentally accessible range of temperatures and irradiation rates, 400 K $< T <$ 1200 K and $10^{-4}$ s$^{-1} < \Phi < 1$ s$^{-1}$. The relaxation volume $V$ which governs the stress dependence of the vacancy energy may be estimated from the structural relaxation of atoms around a vacancy, indicating $V \approx 0.14 A^*d_0$ \cite{nicholson75}. As initial condition we consider a stress- and defect free structure, i.e., $n_i^{\rm I} = n_i^{\rm V} = 0$ and $d_i = d_0$ for all $i$. The CO radius is $R(t)=r_1(t)$ and the initial radius is $R_0 = Nd_0$.  
\begin{figure}[tb]
\begin{minipage}{8cm}
\centerline{
\psfig{figure=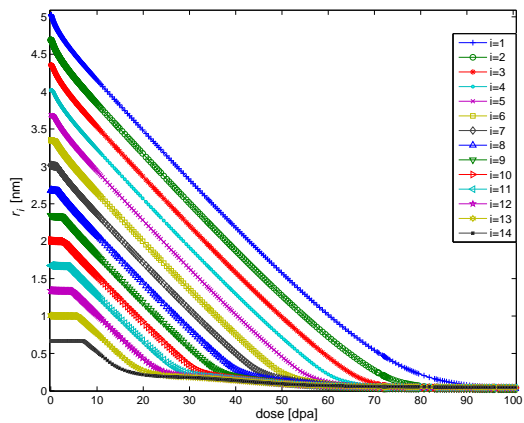,width=8cm,clip=!}
}
\caption{Stage of constant-rate contraction of an irradiated CO, parameters as in Figure \ref{Fig:shell_core}.}
\label{Fig:contraction}
\end{minipage}
\end{figure}
\begin{figure}[tb]
\begin{minipage}{8cm}
\centerline{
\psfig{figure=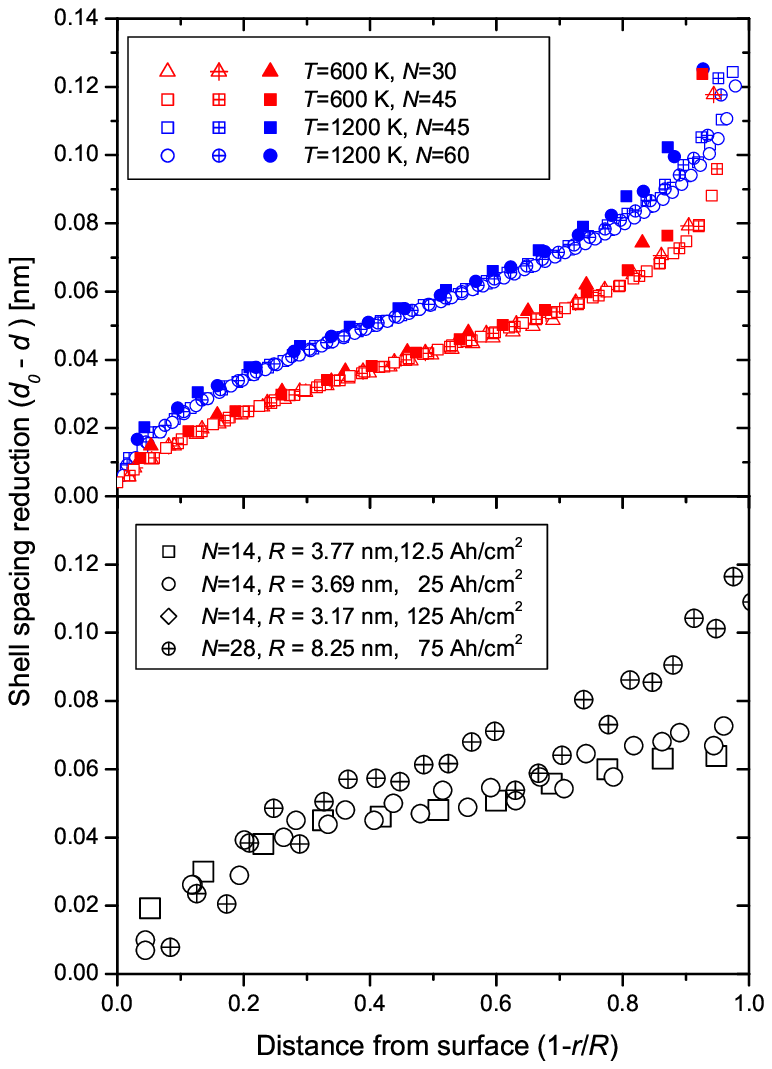,width=8cm,clip=!}
}
\caption{Top: shell spacing profiles during self-similar contraction of CO with different initial shell numbers $N$ and irradiation temperatures $T$. Empty symbols: $R/R_0$ = 0.9, cross center symbols: $R/R_0 = 0.6$, full symbols: $R/R_0 = 0.3$. Bottom: Contraction of a CO with $N=14$ during irradiation at $1000$ K with 1.2 MeV e$^-$ (flux density $150$ A/cm$^2$) \protect\cite{lyutovic00}; the experimental data were taken after irradiation doses of 6dpa, 12dpa and 60dpa; Cross-center symbols: CO with 28 shells after irradiation under the same conditions to approximately 30 dpa.} 
\label{Fig:scaling}
\end{minipage}
\end{figure}

The evolution of an irradiated CO proceeds in two stages, characterized by stress build-up followed by diamond nucleation and growth or, if pressures are insufficient for diamond nucleation, by self-similar contraction of the nanostructure. At the onset of irradiation, quasi-stationary point defect populations build up. Above room temperature these are low everywhere with the exception of ${\cal S}_1$ where loss of atoms by sputtering leads to a gradually increasing excess of vacancies. Displacements of atoms next to these vacancies lead through di-vacancy arrangements to gradual shrinkage of ${\cal S}_1$ which then exerts pressure on ${\cal S}_2$ while itself experiences a tensile hoop stress. These stresses modify the point defect recombination rates: Interstitials created between the two shells now preferentially recombine with vacancies on ${\cal S}_1$. Accordingly, the shrinkage rate of ${\cal S}_1$ is reduced while that of ${\cal S}_2$ is increased as excess vacancies build up. This relieves the compressive stress on ${\cal S}_2$ and transfers it to ${\cal S}_3$ while the spacing between ${\cal S}_1$ and ${\cal S}_2$ henceforth remains approximately constant. The process then repeats itself with ${\cal S}_4$, and so on. As a result, during the stage of stress build-up the onion consists of a set of outer shells that shrink at approximately equal rate. These outer shells are under tensile stress and have high vacancy concentrations. They surround a stress-free non-contracting core with low vacancy concentration. The outer shells are separated from the core by an interface where large compressive stresses reduce the vacancy recombination rate and thus build-up excess vacancies and induce shrinkage. The evolution of the nanostructure is characterized by inwards propagation of the interface and the accompanying pressure wave. The process is illustrated in Figure \ref{Fig:shell_core}. 

The height of the compressive stress peak at the interface between the contracting outer layers and the stationary core increases as the interface propagates inwards. Systematic simulations show that the compressive stress maximum (the highest stress acting on an inner layer) reached during onion contraction is approximately  
\begin{equation}
\sigma_{\rm c,max}\approx f \frac{Nk_{\rm B}T}{V}
\end{equation}
where $f \approx 1.7$. For an onion with $N=10$ deforming at $T=1000$K, compressive stresses may reach up to 27 GPa. The further evolution of the nanostructure depends on whether the peak compressive stress is sufficient for diamond nucleation. If this is not the case, we reach a stage where all shells contract at approximately equal rate proportional to $\Phi$ (Figure \ref{Fig:contraction}) and CO contraction goes along with the progressive 'disappearance' of shells in the onion centre. The shell spacing profile is characterized by a single scaling function $d(r/R)$ which depends only on the ratio $k_{\rm B} T/V$ and remains invariant as $R$ decreases -- we thus speak of self-similar contraction. 

We compare the predictions for this contraction regime with observations of irradiated CO by Banhart and Lyutovic \cite{banhart98,redlich98,lyutovic00}. In these studies, irradiation was carried out in a 1.2 MeV high-voltage electron microscope with typical beam intensities of $4.5 \times 10^{5}$ A/m$^2$, corresponding to typical displacement rates of 10$^-2$ dpa/s. Irradiation at temperatures around 1000K induced pronounced compression after doses of a few dpa \cite{lyutovic00}. Shell radii were determined from electron micrographs by tracing the contrast peak which indicates the location of a shell and determining the enclosed area. Shell spacing profiles obtained at different irradiation doses are shown in Figure \ref{Fig:scaling} for onions with initially 14 and 28 shells (the 28-shell onion is shown in Figure 1). The figure demonstrates that the onions indeed contract in a quasistationary manner with approximately time-independent shell spacings. The calculated shell spacing profiles are in good agreement with the experimental data. 

In CO above a critical size and/or irradiation temperature, the calculated pressures are sufficient to explain the nucleation of diamond in the CO core. Computations indicate a critical stress of about 16 GPa for sp$^3$ bond formation during compression of MWNT and 12 GPa for Graphite\cite{guo04}, which allows us to estimate the 'phase boundary' for diamond nucleation in CO as $NT$[K] > 4500-6000.  Once nucleated, the continuous conversion of sp$^2$ bonded graphene to sp$^3$ bonded diamond maintains the pressure at the shell-core interface at the critical level required for diamond growth. A study of the growth regime needs, however, to take into account the simultaneous operation of other irradiation-driven diamond growth mechanisms \cite{zaiser97} and is thus beyond the scope of the present letter. \\

\end{document}